\documentclass[5p]{elsarticle}
\usepackage{graphicx}
\usepackage{epsfig,latexsym}

\newcommand{\be}{\begin{equation}}
\newcommand{\ee}{\end{equation}}
\newcommand{\ber}{\begin{eqnarray}}
\newcommand{\eer}{\end{eqnarray}}

\begin{document}
\title{The Rise and Fall of the Ridge in Heavy Ion Collisions}


\author[bnl]{P. Sorensen\corref{cor1}}
\ead{prsorensen@bnl.gov}
\author[ens]{B. Bolliet} 
\author[pri]{\'Agnes M\'ocsy}
\author[ksu]{Y. Pandit}
\author[pju]{N Pruthi}

\cortext[cor1]{Corresponding author}

\address[bnl]{Brookhaven National Laboratory, Physics Department, Upton, NY 11973, USA}
\address[ens]{ENS de Lyon, Lyon Cedex 07, France}
\address[pri]{Pratt Institute, Department of Math and Science, Brooklyn, NY 11205, USA}
\address[ksu]{Kent State University, Physics Department, Kent, OH 44242, USA}
\address[pju]{Panjab University, Physics Department, Chandigarh 160014, India }

\begin{abstract}
  Recent data from heavy ion collisions at RHIC show unexpectedly
  large near-angle correlations that broaden longitudinally with
  increasing centrality. The amplitude of this ridge-like correlation
  rises rapidly, reaches a maximum, and then falls in
  the most central collisions. In this letter we explain how this
  behavior can be uniquely explained by initial-state coordinate-space
  anisotropies converted into final-state momentum-space correlations. We
  propose $v_n^2/\varepsilon_{n,\mathrm{part}}^{2}$ as a useful way to study
  length scales and provide a prediction for the ridge in Pb+Pb
  collisions at $\sqrt{s_{\mathrm{NN}}}=$ 2.76 TeV.
\end{abstract}

\begin{keyword}
heavy ion collisions \sep correlations \sep flow \sep ridge
\PACS
\end{keyword}

\maketitle

\section{Introduction}

The motivation for colliding heavy ions at facilities like the Relativistic Heavy Ion Collider (RHIC) at Brookhaven National Laboratory and the Large Hadron Collider (LHC) at CERN is to form a state of matter called the Quark Gluon Plasma (QGP) \cite{Reisdorf:1997fx}. Each of these collisions deposit many TeV of energy into a region roughly the size of the colliding nuclei. That region is so hot and dense that quarks and gluons become the relevant degrees of freedom instead of hadrons~\cite{eos}. This QGP existed when the universe was less than a microsecond old, and still extremely hot and dense.

Two-particle correlations measured at RHIC reveal features unique to Nucleus-Nucleus collisions~\cite{onset,ridgedata}. Pair densities are commonly measured as a function of the azimuthal angle and pseudo-rapidity difference between the particles ($\Delta\phi$ and $\Delta\eta$ respectively). While two-particle correlations in p+p and d+Au collisions show a narrow peak centered at $\Delta\phi$ and $\Delta\eta=0$, the near-angle peak in Au+Au collisions broadens longitudinally and narrows in azimuth relative to p+p collisions. An analysis of the width of the peak for particles of all transverse momentum $p_T$ finds the correlation extends across nearly two units of pseudo-rapidity~\cite{ridgedata}. When selecting higher $p_T$ particles ({\it e.g.} $p_T>2$~GeV/c), the correlation extends beyond the acceptance of the STAR detector ($\Delta\eta<2$) and perhaps as far as $\Delta\eta=4$ as indicated by PHOBOS data~\cite{ridgedata}. This feature is known as the ridge.

STAR data shows that the ridge amplitude rises rapidly with collision centrality~\cite{onset} before reaching a maximum and falling off in the most central bins. The drop is often ignored. In this letter we present an explanation for the centrality dependence of the ridge amplitude related to density inhomogeneities in the initial overlap region. We use measurements of the second harmonic momentum-space anisotropy $v_2$ and single particle rapidity density $dN/dy$ along with a Monte-Carlo Glauber model~\cite{glauber} for the initial density to predict the amplitude ($A_1$) of the near-side ridge correlation as a function of centrality for $\sqrt{s_{\mathrm{NN}}}=62.4$ GeV, 200 GeV, and 2.76 TeV. A successful description of $A_1$ is noteworthy, because it demonstrates that the correlations measured in heavy-ion collisions represent an image of structures such as flux-tubes~\cite{Voloshin:2003ud,mclerran} in the initial overlap of heavy ion collisions of the order of $10^{-15}$ m in size.

In Glauber Monte-Carlo models~\cite{glauber} the density distribution in the initial collision region is assumed to be determined by the positions of the nucleons participating in the interactions (participants). For a typical collision, the shape of the overlap region will be predominantly an ellipse. The eccentricity of the ellipse can be quantified as $\varepsilon_{\mathrm{std}}=\frac{\langle y^2\rangle-\langle x^2\rangle}{\langle y^2\rangle+\langle x^2\rangle}$. Interactions amongsts the systems constituents can convert that elliptic shape from coordinate-space into momentum-space leading to a large second Fourier component in the momentum-space distribution \textit{i.e.} $v_2 = \langle\cos2(\phi-\Psi_{\mathrm{RP}})\rangle > 0$, where $\phi$ is the azimuth angle of particles emitted from the collision and $\Psi_{\mathrm{RP}}$ is the reaction plane angle defined by the vector connecting the centers of the two colliding nuclei. There are a finite number of participants in each collision so there will be event-to-event fluctuations in the density distributions that lead to fluctuations in the initial eccentricity as well as a tilt of the elliptic shape of the overlap region away from $\Psi_{\mathrm{RP}}$. The plane containing the titled axis and beam axis is called participant plane, and the eccentricity calculated relative to that axis is called the participant eccentricity $\varepsilon_{\mathrm{part}}$~\cite{participant}. $\varepsilon_{\mathrm{part}}$ is a positive definite quantity so it can lead to non-zero $v_2$ even for head-on collisions. Fluctuations in $\varepsilon_{\mathrm{part}}$ can lead to fluctuations in $v_2$~\cite{v2fluct}. It was argued previously that eccentricity fluctuations can give rise to $v_n$ fluctuations for more harmonics than just $n=2$~\cite{mishra}, and that those fluctuations could therefore be the source of the near-side ridge~\cite{sorensen1} (especially if the $v_n$ fluctuations depend on $\Delta\eta$ \textit{i.e.} $\langle v_n(\eta)v_n(\eta+\Delta\eta)\rangle \equiv f(\Delta\eta)$). Subsequent calculations from several groups~\cite{mclerran,v3,brazil,hannah} support this. To test this conjecture, we calculate the centrality dependence of the near-side ridge amplitude $A_1$ from eccentricity fluctuations.

\section{Eccentricity Fluctuations, Length Scales, and The Ridge}
 
We base our calculations on three premises: 1) the expansion of the fireball in heavy-ion collisions converts anisotropies from coordinate-space into momentum-space, 2) the conversion efficiency increases with density, and 3) the relevant expansion plane is the participant plane. The participant plane can be defined for any harmonic number. A system with a lumpy initial energy density will lead to finite participant eccentricity at several harmonics~\cite{v3,teaneyv3}; eccentricity can be thought of as a harmonic decomposition of the azimuthal dependence of the initial density. To illustrate how eccentricity fluctuations can lead to a ridge-like structure in particle correltions, using the definition in Ref.~\cite{v3}, we calculate $\varepsilon_{n,\mathrm{part}}^2=\frac{\langle r^2\cos(n\phi)\rangle^2+\langle r^2\sin(n\phi)\rangle^2}{\langle r^2\rangle}$ for all harmonics.

\begin{figure*}[htb]
 \centering
 \resizebox{0.32\textwidth}{!}{\includegraphics{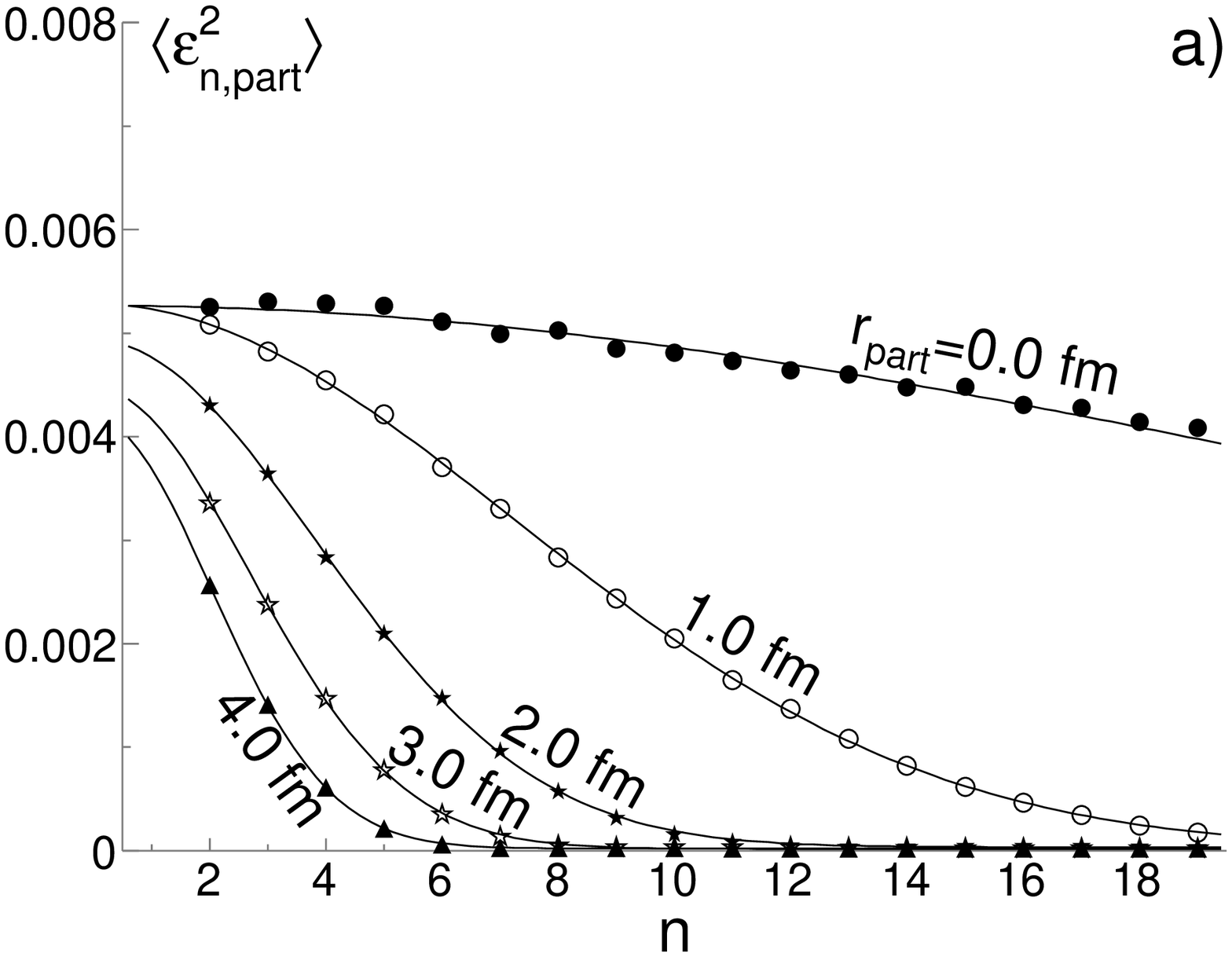}}
 \resizebox{0.32\textwidth}{!}{\includegraphics{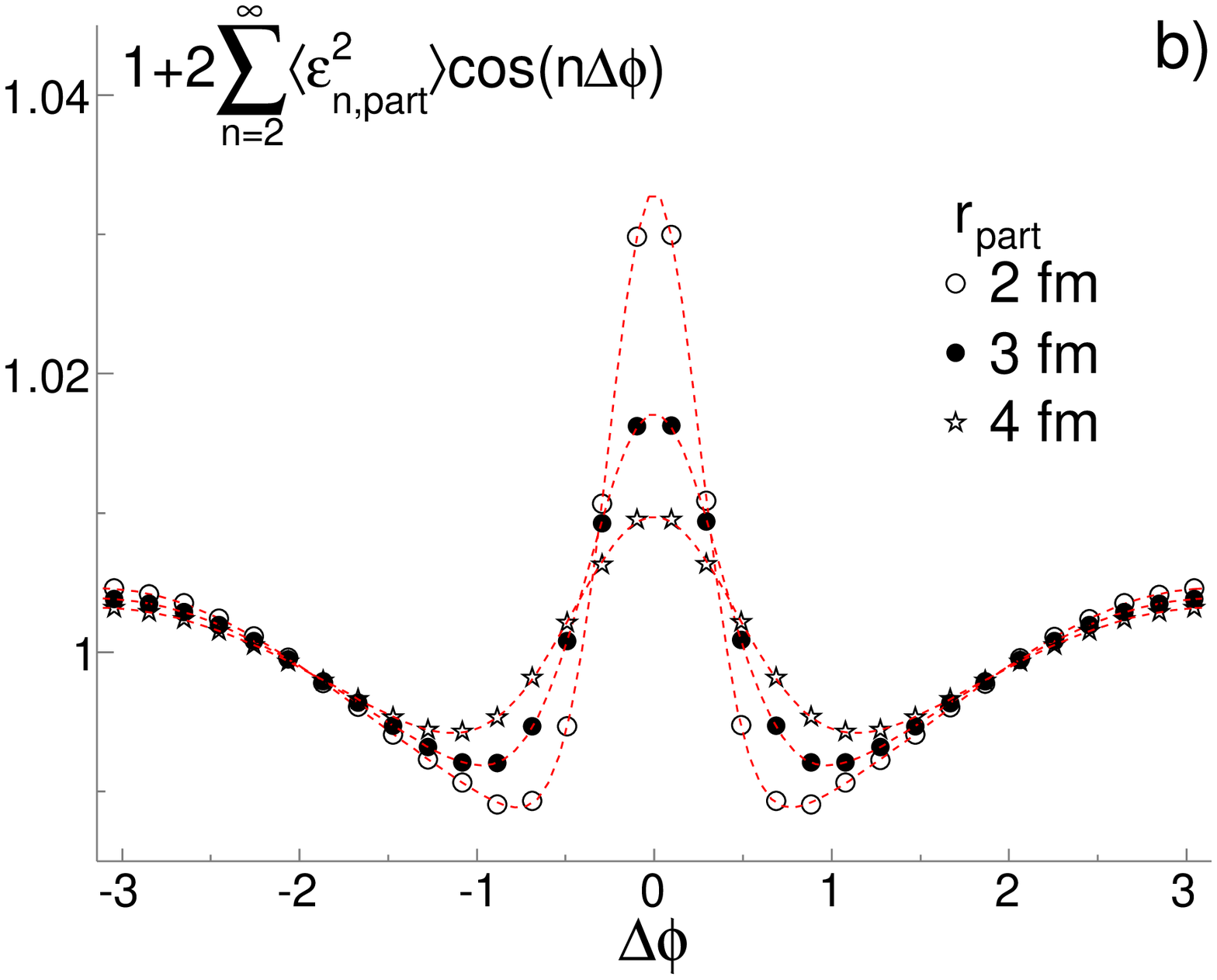}}
 \resizebox{0.32\textwidth}{!}{\includegraphics{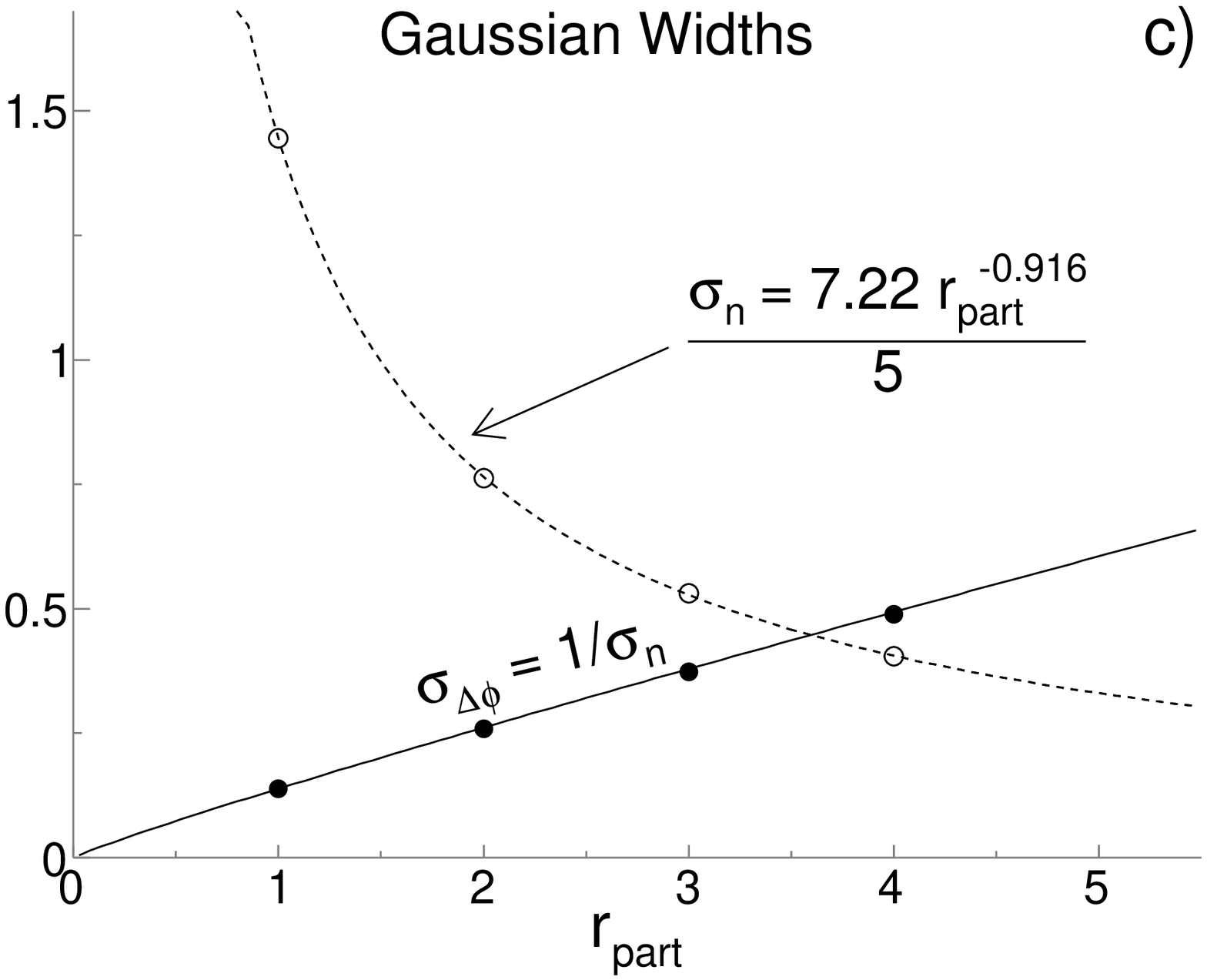}}
 \caption[]{ a) $\langle\varepsilon_{n,\mathrm{part}}^2\rangle$ for central
   Au+Au collisions from a Glauber Monte-Carlo where participants are
   treated as point-like or smeared over a region of size
   $r_{\mathrm{part}}$. The curves show Gaussian fit functions. b) the
   $\Delta\phi$ dependence of two-particle correlations if $\langle
   v_n^2\rangle \propto \langle\varepsilon_{n,\mathrm{part}}^2\rangle$. The
   curves show a Gaussian+$\cos(\Delta\phi)$ fitting function. Bottom
   panel: The Gaussian widths extracted from the fits in panel a) and
   b) with a power-law fit to describe the $r_{\mathrm{part}}$ dependence. }
\label{f1}
\end{figure*}

Fig.~\ref{f1} (a) shows the $n^{th}$-harmonic participant eccentricity $\langle\varepsilon_{n,\mathrm{part}}^2\rangle$ for central Au+Au collisions from our Monte-Carlo Glauber model. Typically the participant eccentricity is calculated based on the positions of point-like participants ({\it i.e.} the participant is said to exist at a precise $x$ and $y$). One can also calculate the eccentricity from a more realistic model with participants smeared over some region. This is done by treating each participant as many points distributed within a disk of radius $r_{\mathrm{part}}$. The figure shows $\langle\varepsilon_{n,\mathrm{part}}^2\rangle$ for $r_{\mathrm{part}}=0,1,2,3,$ and 4 fm. Increasing $r_{\mathrm{part}}$ washes out the higher $\langle\varepsilon_{n,\mathrm{part}}^2\rangle$ terms. The dependence of $\langle\varepsilon_{n,\mathrm{part}}^2\rangle$ on $n$ is well fit with a Gaussian $e^{-\frac{1}{2}(\frac{n}{\sigma_n})^2}$ for all values of $r_{\mathrm{part}}$ with the width of the Gaussian $\sigma_{n}$ narrowing as $r_{\mathrm{part}}$ is increased.

Here we've introduced the length scale $r_{\mathrm{part}}$ causing the higher terms in $\langle\varepsilon_{n,\mathrm{part}}^2\rangle$ to be washed out. That effect is more general though and we believe it is important for understanding correlations and $v_n$ fluctuations~\cite{sound}. One can also consider what happens when particles free-stream for some amount of time $\tau_{\mathrm{fs}}$ before they interact; which also introduces a length scale $c\tau_{\mathrm{fs}}$ leading to a reduction of higher terms~\cite{hannah}. The mean-free-path ($l_{\mathrm{mfp}}$) a particle travels between interactions in the fireball also affects the ability of the fireball to convert higher $\langle\varepsilon_{n,\mathrm{part}}^2\rangle$ terms into $v_n^2$~\cite{sound,ollialver}. The $l_{\mathrm{mfp}}$ will prevent higher $\langle\varepsilon_{n,\mathrm{part}}^2\rangle$ terms from being converted into $v_n^2$. The acoustic horizon may also play an important role~\cite{sound}. In our calculations, we study the effect of varying $r_{\mathrm{part}}$ with the understanding that many length-scales, like $l_{\mathrm{mfp}}$ and $c\tau_{\mathrm{fs}}$, will contribute to the final dependence of $v_n$ on $n$.

$v_n^2$ is related to the two-particle correlations $dN_{pair}/d\Delta\phi$ by a Fourier transform. If $v_n^2$ {\it vs.} $n$ has a Gaussian shape then $dN_{pair}/d\Delta\phi$ will also have a Gaussian peak at $\Delta\phi=0$. Fig.~\ref{f1} (a) shows that $\langle\varepsilon_{n,\mathrm{part}}^2\rangle$ follows a Gaussian so that eccentricity fluctuations should lead to a near-side Gaussian with a width depending on the length scales in the system. Fig~\ref{f1} (b) shows the shape of the two-particle correlations from $\langle\varepsilon_{n,\mathrm{part}}^2\rangle$. For $n=1$, $\langle\varepsilon_{n,\mathrm{part}}^2\rangle=0$ since $\langle\varepsilon_{n,\mathrm{part}}^2\rangle$ is calculated in the center-of-mass frame of the participants. This leads to a  Gaussian centered at $\Delta\phi=0$ and an apparent negative $\cos(\Delta\phi)$ term due to the suppression of the $n=1$ eccentricity fluctuations. The negative $\cos(\Delta\phi)$ is seen in the data to follow the same centrality dependence as the nearside Gaussian~\cite{onset}. The fact provides evidence that both the nearside Gaussian and the awayside negative $\cos(\Delta\phi)$ terms are related to initial density fluctuations.

We fit the correlations in panel (b) with a Gaussian of width $\sigma_{\Delta\phi}$ and a $\cos(\Delta\phi)$ term. In Fig.~\ref{f1} (c) we plot $\sigma_{n}$ (scaled by 1/5 for clarity) and $\sigma_{\Delta\phi}$.  $\sigma_{n}(r_{\mathrm{part}})$ is fit with a power law $\sigma_{n}(r_{\mathrm{part}})=7.22r_{\mathrm{part}}^{-0.916}$ and $\sigma_{\Delta\phi}$ is described the inverse. Increasing length scales in the system should lead to a broadening of the azimuthal width of the near-side Gaussian. Having demonstrated that it is natural for eccentricity fluctuations to lead to a ridge like correlation with a width dependent on the length scales in the system, we now consider the amplitude of the ridge.

\section{The Ridge Amplitude}
 
\begin{figure}[htb]
 \centering
 \resizebox{0.4\textwidth}{!}{\includegraphics{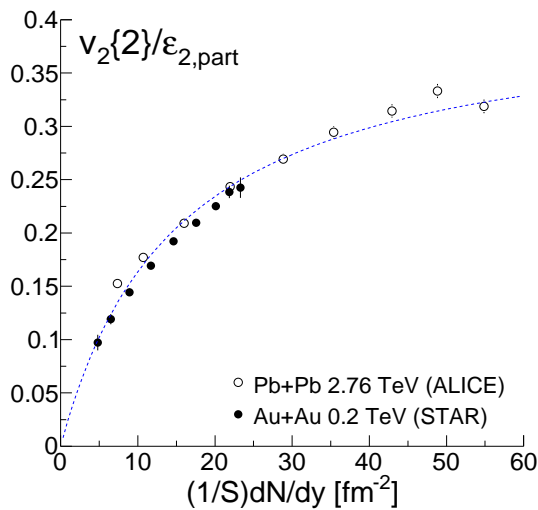}}
 \caption[]{ The ratio of $v_2\{2\}$ over $\varepsilon_{2,\mathrm{part}}$ from
   a Glauber Monte-Carlo {\it vs.} $(1/S)dN/dy$ for $\sqrt{s_{_{\mathrm{NN}}}}=0.2$
   and 2.76 TeV. The fit function
   $v_2\{2\}/\varepsilon_{2,\mathrm{part}}=A/(1+\frac{B}{(1/S)dN/dy})$ is
   adapted from Ref.~\cite{knudsen}. The fitting parameters are
   $A=0.412$ and $B=15.175$. }
\label{f2}
\end{figure}

The ridge amplitude $A_1$ is found from data by fitting $\Delta\rho/\sqrt{\rho_{\mathrm{ref}}}$ (the pair density $\rho$ minus the reference pair density $\rho_{\mathrm{ref}}$ scaled by $\sqrt{\rho_{\mathrm{ref}}}$) {\it vs.} $\Delta\phi$ and $\Delta\eta$~\cite{daugherity}. The fit function has $\Delta\eta$ independent $\cos(\Delta\phi)$ and $\cos(2\Delta\phi)$ terms, and a near-side 2-D Gaussian with amplitude $A_1$. We work with the conjecture that the 2-D Gaussian is a manifestation of $\langle\varepsilon_{n,\mathrm{part}}^2\rangle$ and calculate the centrality dependence of $A_1$. Our result for $A_1$ will be related to $v_n^2$ so we need to know the transfer function or conversion efficiency $c_n$~\cite{sound} of $\langle\varepsilon_{n,\mathrm{part}}^2\rangle$ into $v_n^2$. We expect $c_n$ to depend on particle density. In Fig.~\ref{f2} we show the two-particle cumulant scaled by eccentricity $v_2\{2\}/\varepsilon_{2,\mathrm{part}}$ {\it vs.} density $(1/S)dN/dy$ from RHIC and LHC~\cite{v2papers,mult}. The funciton from Ref.~\cite{knudsen} fits the full range of RHIC and LHC data reasonably well providing our estimate of $c_2$.

We can estimate $A_1$ independently from the fluctuations of several different harmonics. Since a $\cos(2\Delta\phi)$ term proportional to $\varepsilon_{\mathrm{std}}^2$ is used in the fit to $\Delta\rho/\sqrt{\rho_{\mathrm{ref}}}$~\cite{david} we base our estimate of $n=2$ component of $A_1$ on the difference between $\varepsilon_{2,\mathrm{part}}^2$ and $\varepsilon_{\mathrm{std}}^2$. We can also calculate the $n=3$ component from $\varepsilon_{3,\mathrm{part}}^2$ alone but this requires an assumption about $c_3$ which is not yet measured.  We need to convert our prediction for correlations from a particular harmonic into a Gaussian amplitude $A_1$. For $n=2$ we find
\begin{equation}
\frac{1}{2\pi}\int_{-\pi}^{\pi}\frac{\Delta\rho}{\sqrt{\rho_{\mathrm{ref}}}}\cos(2\Delta\phi)d\Delta\phi
= 0.11A_1
\end{equation}
with the $n=3$ component similary giving $0.039A_1$. The azimuthal width $\sigma_{\Delta\phi}$ of the near-side Gaussian is weakly dependent on centrality so we use a typical value of $\sigma_{\Delta\phi}=0.65$.
To relate $v_n^2$ to $\Delta\rho/\sqrt{\rho_{\mathrm{ref}}}$ we need to include the particle density $\rho_0=\frac{1}{2\pi}\frac{dN}{dy}$~\cite{rho0}. Combining $\langle\varepsilon_{2,\mathrm{part}}^2\rangle-\langle\varepsilon_{\mathrm{std}}^2\rangle$ with the conversion efficiency $c_2$, particle density $\rho_0$, and factor of 0.11 we find $A_1 \approx \rho_0c_2(\varepsilon_{2,\mathrm{part}}^2-\varepsilon_{\mathrm{std}}^2)/0.11$.  We take $\rho_0$ from data, $c_2$ from Fig.~\ref{f2}, and $\langle\varepsilon_{2,\mathrm{part}}^2\rangle$ and $\varepsilon_{\mathrm{std}}$ from our Monte-Carlo Glauber model. The $n=3$ based prediction does not need $\varepsilon_{\mathrm{std}}$ but does require the conversion efficiency $c_3$. We assume that $c_3$ has the same density dependence as $c_2$ but that various length scales in the system will suppress $c_3$ relative to $c_2$.  $c_3 = c_2/2.6$ gives good agreement between the $n=2$ and $n=3$ estimates and is consistent with the AMPT results in Ref.~\cite{v3}. From $n=3$ we find $A_1 \approx \rho_0c_3\varepsilon_{3,\mathrm{part}}^2/0.039$ where $c_3\approx c_2/2.6$.

\begin{figure}[htb]
\centering
\resizebox{0.4\textwidth}{!}{\includegraphics{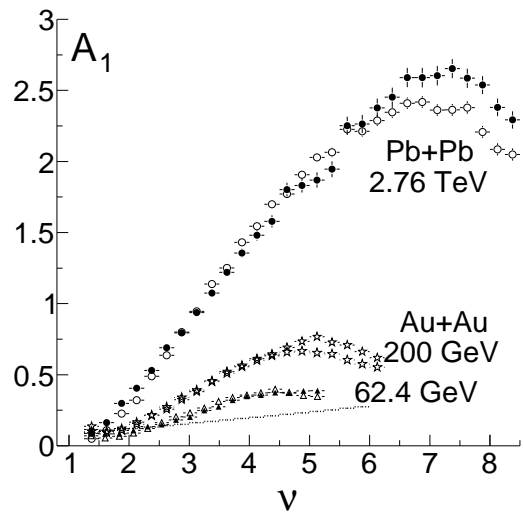}}
\caption[]{ An estimate of the contribution of eccentricity
 fluctuations to the near-side Gaussian peak amplitude
 $A_1$ {\it vs.} centrality measure $\nu=2N_{bin}/N_{\mathrm{part}}$ for 62.4 Au+Au,
 200 GeV Au+Au, and 2.76 TeV Pb+Pb collisions. The closed symbols
 are the amplitudes expected from  $\varepsilon_{2,\mathrm{part}}^2 - \varepsilon_{\mathrm{std}}^2$ and the
 closed symbols are from $\varepsilon_{3,\mathrm{part}}^2$. }
\label{f3}
\end{figure}

Fig.~\ref{f3} shows our estimate of the ridge amplitude $A_1$ based on $\langle\varepsilon_{n,\mathrm{part}}^2\rangle$ {\it vs.} centrality parameter $\nu=2N_{\mathrm{bin}}/N_{\mathrm{part}}$ for Au+Au collisions at 62.4 and 200 GeV and for Pb+Pb collisions at 2.76 TeV. $N_{\mathrm{bin}}$ and $N_{\mathrm{part}}$ are the number of binary nucleon-nucleon collisions and the number of participants. More central or higher energy collisions yield larger values of $\nu$. The open (closed) symbols show the $n=3$ ($n=2$) estimates.

\section{The Rise and Fall}
 
\begin{figure}[htb]
\centering
\resizebox{0.48\textwidth}{!}{\includegraphics{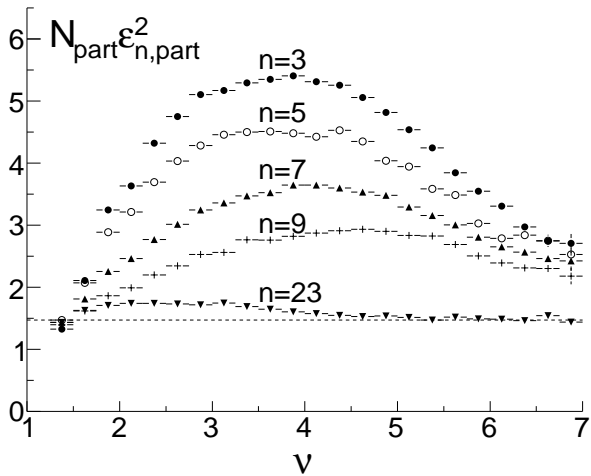}}
\caption[]{ The centrality dependence of odd terms for $N_{part}\varepsilon_{n,\mathrm{part}}^2$.  
The rise and fall of the ridge can be traced to the non-statistical centrality dependence of $\varepsilon_{n,\mathrm{part}}^2$ caused by the intrinsic almond shape of the nuclear overlap region. }
\label{f4}
\end{figure}

We find that our estimate of $A_1$ agrees with what has been observed at 200 and 62.4 GeV.  Our $A_1$, like the data, starts at a small value and rises much faster than expectations from a linear superposition of independent p+p collisions (shown for 200 GeV in the Fig.~\ref{f3}) which assumes that correlations grow as $N_{bin}/dN/dy$. The rise continues until $A_1$ reaches a maximum of 0.7 for 200 GeV near $\nu=5$, then falls again.  The 2.76 TeV calculations show a similar trend as the 200 and 62.4 GeV calculations but the amplitude is expected to be much larger. This provides a testable prediction for the LHC experiments.
 Our picture provides a natural explanation for the rise and fall related to the initial overlap geometry and its fluctuations.  For the $n=3$ term for example, $\langle\varepsilon_{3,\mathrm{part}}^2\rangle$ falls with $N_{\mathrm{part}}$ since the larger $N_{\mathrm{part}}$ leads to smaller fluctuations. But $N_{\mathrm{part}}\langle\varepsilon_{3,\mathrm{part}}^2\rangle$ first rises then falls (see Fig.~\ref{f4}). This rise and fall is due to the almond shaped geometry of the overlap region which leads to non-statistical fluctuations in $\varepsilon_{n,part}^2$ for more than just the $n=2$ harmonic~\cite{teaneyv3}.  Fig.~\ref{f4} shows $N_{part}\varepsilon_{n,\mathrm{part}}^2$ for $n=3$, 5, 7, 9, and 23. For harmonics close to $n=2$, the almond shape of the overlap zone causes large deviations from a trivial $\varepsilon_{n,\mathrm{part}}^2 \propto 1/N_{part}$ behavior. The higher harmonics approach this statistical expectation with $n=23$, nearly reaching a $1/N_{part}$ behavior.

Both $c_n$ and $\rho_0$ are increasing with centrality but the product of $\rho_0c_n\langle\varepsilon_{n,\mathrm{part}}^2\rangle$ rises until very central collisions and then falls. The drop fall in central is due to the subsidence of the effect of the almond shape of the overlap geometry. The observation that the rise and fall exists in the near-side ridge amplitude shows that the ridge is dominated by initial geometry fluctuations; even exhibiting a dependence on the coupling of various harmonics to the intrinsic almond shape of the overlap region. We know of no other plausible scenarios to explain the rise and fall of the ridge other than this explanation. This rise and fall is to our knowledge a unique signature of density fluctuations. This implies that momentum-space correlations are sensitive to initial density fluctuations and that correlations in heavy-ion collisions provide an image of the density distributions in the initial overlap zone. The system created in these collisions acts as a femtoscope, revealing structures with length scales on the order of a fm.

\section{Discussion and Conclusion}

Our explanation for the centrality dependence of $A_1$ provides a natural explanation for the rise and fall of the ridge. Our estimates of the amplitude agree with RHIC data suggesting that the conversion of geometry fluctuations in the initial overlap region into momentum space causes the near-side ridge structure. We expect the conversion efficiency to drop with $n$ since effects like initial-state free-streaming and mean-free-path will wash out the higher harmonic terms. Extracting the conversion efficiency $c_n=\frac{v_n\{2\}^2}{\varepsilon_{n,\mathrm{part}}\{2\}^2}$ as a function of $n$~\cite{sound}, centrality, and particle kinematics will provide information on those effects. This only relies on measuring the two-particle cumulant $v_n\{2\}^2$ which is a rather simple experimental measurement and comparing it to the initial eccentricities from various models of the initial density. It will be particularly interesting to determine $c_n$ as a function of $\Delta\eta$ to understand how de-coherence affects manifest in the longitudinal direction.

We presented the participant eccentricity {\it vs.} harmonic when the participants are treated as point-like or smeared over a radius $r_{\mathrm{part}}$. The larger values of $r_{\mathrm{part}}$ wash out the higher harmonic eccentricities. We argued that, similarly, a large mean-free-path or acoustic horizon should also wash out higher harmonics of $v_n$. Such an effect could lead to a Gaussian peak in two-particle correlations at $\Delta\phi=0$ similar to that seen in the data.  We calculated the contribution to the near-side Gaussian peak that we expect from initial density fluctuations. Following simple premises, we find that the near-side peak from density fluctuations should rise rapidly, reach a maximum just before the most central events, then fall. Our estimate of the magnitude is in agreement within our uncertainties with the available data and the shape matches the data. This is the only calculation we know of to correctly describe the rise and fall of the ridge amplitude. Our calculation shows that the rise and fall is related to the interaction of fluctuations with the shape of the initial overlap geometry. We conclude therefore that density fluctuations are the dominant source for the ridge-like correlations. The longitudinal width of the ridge remains an interesting topic to investigate. Finally we use the same framework to predict the ridge amplitude for 2.76 TeV Pb+Pb collisions.

The authors thank Sergei Voloshin, Sean Gavin and Joern Putschke for their helpful comments.

\end{document}